\documentclass[twocolumn,showpacs,preprintnumbers,amsmath,amssymb]{revtex4}
\usepackage{graphicx}
\usepackage{dcolumn}
\usepackage{bm}

\begin{document}
\draft

\title{Kondo hole in one-dimensional Kondo insulators}
\author{Isao Maruyama$^{1}$, Naokazu Shibata$^{2}$ and Kazuo Ueda$^{1,3}$}
\address{
$^{1}$Institute for Solid State Physics, University of Tokyo,
5-1-5 Kashiwa-no-ha, Kashiwa, Chiba 277-8581, Japan\\
$^{2}$Department of Basic Science, University of Tokyo
3-8-1 Komaba, Meguro, Tokyo 153-8902, Japan\\
$^{3}$Advanced Science Research Center, Japan Atomic Energy Research Institute,
Tokai, Ibaraki 319-1195, Japan}
\date{\today}

\begin{abstract}
Properties of a non-magnetic impurity in Kondo insulators
are investigated
by considering a one-dimensional Kondo lattice model
with depletion of a localized spin.
The ground state phase diagram determined by the Lanczos method
shows
that the magnetic moment is more stable than in ordinary metals.
Temperature dependence of impurity susceptibilities
is also studied by using
the finite temperature density-matrix renormalization
group.
\end{abstract}

\maketitle

\section{Introduction}
The single-impurity Anderson model~\cite{ANDERSON}
was introduced
to study the condition for
formation of localized
moments in metals.
The self-consistent Hartree-Fock method
showed that there is a transition between the magnetic state
and the non-magnetic state at $U_c \sim 1/\rho_d$,
where $U$ is on-site Coulomb energy of a virtual bound state,
and $\rho_d$ is its density of states at the Fermi level.
But as a result of  the Kondo effect~\cite{KONDO},
the sharp transition disappears
and becomes a cross-over.
That is,
at high temperatures higher than the Kondo temperature,
there is the region
where thermodynamic properties are well described by the localized
magnetic moment.
At zero temperature,
however,
the magnetic moment disappears
by the Kondo effect leading to  
the singlet ground state
which is obtained by K.Yosida~\cite{YOSIDA}.

In metals,
gap-less excitations around the Fermi energy
play an important role to the Kondo effect.
On the contrary,
in spin-1/2 anti-ferromagnetic (AF) Heisenberg two-leg ladders,
which have  spin gaps and short range singlet correlations,
a non-magnetic impurity,
{\it i.e.} depletion of a spin,
induces a local magnetic moment
around the vacant site
because
the impurity breaks a dimer singlet pair of
the resonating valence bond (RVB) state
and leaves an unpaired spin~\cite{FUKUYAMA,MOTOME}.
In fact,
the magnetic moments exhibit a Curie law susceptibility,
whose Curie constant is proportional to the concentration of
the impurities~\cite{IINO}.
Correlations between
the magnetic moments are staggered
and
rapidly decaying with the impurity distance.
Experimentally,
when concentration of the impurities
exceeds a critical value,
the spin gap disappears
and
it leads to an instability of the spin-liquid state towards an AF
ordered state~\cite{AZUMA}.
On the other hand,
theoretically,
the critical concentration
seems to be zero
for a bipartite lattice
and thus any finite concentration of
impurities leads to the ground state
with AF quasi long range order and gap-less excitations.

An impurity in the Kondo Lattice (KL) model
which describes heavy fermion system
is introduced as a depletion of a localized spin
similarly to the impurities in the two-leg spin ladder.
However,
since 
the KL model includes both itinerant electrons and localized spins
in contrast to the spin ladder systems,
after the depletion
one should ask necessary condition for formation of the localized
moment on the impurity site.
The model Hamiltonian for general dimension $d$ is defined as
\begin{eqnarray}
H&=&\sum_{\langle n,n'\rangle ,s} t_{n,n'} c^\dagger_{ns}c_{n's} 
+J\sum_{n\neq n_d} {\bf S}_n \cdot {\bf s}_n
\nonumber\\&&
+\epsilon \sum_{s}c^\dagger_{n_d s}c_{n_d s}
+Uc^\dagger_{n_d \uparrow}c_{n_d \uparrow}c^\dagger_{n_d \downarrow}c_{n_d \downarrow}
\label{eqHm}
,
\end{eqnarray}
where
${\bf s}_n:= \sum_{s,s'} {1\over 2}
{\bf \sigma}_{s,s'}c^\dagger_{ns}c_{ns'}$
and
$n$ is $d$-dimensional vector
and
$n_d$ indicates the impurity site.
This model represents a single impurity and
the surrounding KL model.
It can be viewed also as
the KL model without the spin at $n=n_d $:
the KL model with a Kondo hole.
When the KL model at half filling,
so-called Kondo insulator,
is considered
we may expect that
the existence of a spin gap and a charge gap
will lead to
similar properties as
the two-leg spin ladder in certain situations.
In fact,
by substituting non-magnetic Lu-ions into a Kondo insulator,
Yb$_{1-x}$Lu$_x$B$_{12}$,
Curie-like susceptibilities are observed at low temperatures
for small concentrations~\cite{IGA}.
This fact suggests that localized magnetic moments are induced
by the substitution of the non-magnetic Lu-ions.
Correspondingly,
the optical gap of the Kondo insulator YbB$_{12}$,
is partially filled by the substitution~\cite{OKAMURA}.

Formation of impurity bands in Kondo insulators is discussed
by Schlottmann and his coworker
by using the periodic Anderson model
by employing the $1/d$ expansion for the self-energy of the host sites
for which the second order perturbation is used~\cite{SCHLOTTMANN}.
Concerning the impurity site,
Coulomb interaction is neglected,
corresponding to $\epsilon=U=0$ in our notation.

\begin{figure}
\includegraphics{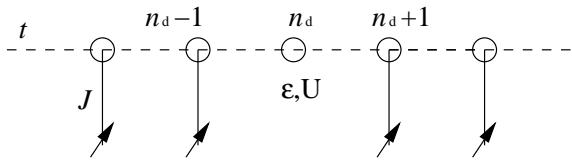}
\caption{The one-dimensional Kondo lattice model
with an impurity site,
$n_d$.}
\label{KLImp}
\end{figure}

In this paper,
we study
the condition for the presence or absence
of localized moments in Kondo insulators
and properties of these moments.
For this purpose,
we will restrict ourselves to the one-dimensional (1D) case
(Fig.~\ref{KLImp}),
setting the hopping parameter $t_{n,n'}=-t=-1$
({\it i.e.}, measuring all energies in units of $t$).
In the one-dimension
we can use various powerful methods like
the numerical exact diagonalization and
the density matrix renormalization group.
Various properties of the pure 1D Kondo insulator:
\begin{eqnarray}
H_{{\rm KL}}=\sum_{\langle n,n'\rangle ,s}t_{n,n'} c^\dagger_{ns}c_{n's} 
+J\sum_{n} {\bf S}_n \cdot {\bf s}_n
\label{eqHKL}
\end{eqnarray}
have been studied extensively~\cite{UEDA}.
In particular,
magnitudes of the charge gap and the spin gap
are obtained numerically~\cite{SHIBATA}.
The spin-spin correlation functions decay with oscillations,
the RKKY oscillations,
and its correlation length $\xi$ are also determined~\cite{RKKYint}.

\section{Property of the ground state}
In order to determine the magnetic or non-magnetic ground state,
the total number of conduction electrons $N_{\rm tot}$
and the total spin $S^z_{\rm tot}$ are studied.
Since the surrounding Kondo insulator has
the totally singlet ground state
when the impurity site is neglected,
we expect that there is one-to-one correspondence between 
$N_{\rm tot}$ and the existence or absence of the magnetic moment
around the impurity site.
Let us start from the strong coupling limit,
$J/t \gg 1$. To make the following discussion clear
we assume  a finite $J$ and $t=0$.
Then the impurity site is independent from the rest
and it is sufficient to 
think
only four states for the impurity site:
the unoccupied impurity state ($S^z_{\rm tot}=0 , N_{\rm tot}=L-1$),
the doublet ($S^z_{\rm tot}=\pm {1\over 2}, N_{\rm tot}=L$),
and the doubly occupied state ($S^z_{\rm tot}=0 , N_{\rm tot}=L+1$),
where $L$ is the number of the lattice sites including
the impurity site.
In this case,
it is clear that
these states are distinguished
with
the total number of conduction electrons of the ground state
$\langle N_{\rm tot}\rangle _{\rm gs}$.
For a finite $t$ various hopping processes take place.  
However, these processes keep the number of conduction electrons
and total spin of the system. 
Since the quantum numbers mentioned above
define a disjoint subspace,
we can investigate the problem of formation of magnetic moment just
by looking at the total number of electrons at finite $J/t$.
The quantum number of the ground state is
obtained by comparing the lowest eigenvalues
with the Lanczos method for different sub-matrices.
We place 
the impurity site at the center
and use the open boundary conditions.

There is an electron-hole symmetry for the present system.
By using the transformation,
$c_{ns} \rightarrow (-1)^n c_{ns}^\dagger $
and
${\bf S}_n \rightarrow -{\bf S}_n$,
the Hamiltonian (Eq.~\ref{eqHm}) is reduced to
\begin{eqnarray}
H'&=&\sum_{\langle n,n'\rangle ,s} t_{n,n'} c^\dagger_{ns}c_{n's} 
+J\sum_{n\neq n_d} {\bf S}_n \cdot {\bf s}_n
\nonumber\\&&
-(\epsilon+U) \sum_{s}c^\dagger_{n_d s}c_{n_d s}
+Uc^\dagger_{n_d \uparrow}c_{n_d \uparrow}c^\dagger_{n_d \downarrow}c_{n_d \downarrow}
\nonumber\\&&
+2\epsilon +U
,
\end{eqnarray}
where the last term $2\epsilon+U$ can be ignored.
Therefore
the physical properties for a set of parameters $(U , \epsilon)$
are identical to the system with $(U, -\epsilon-U)$.
In particular,
since $n_{ns}$ goes to $1-n_{ns}$
under the transformation,
this symmetry leads to
\begin{eqnarray}
\langle N_{\rm tot} \rangle_{\rm gs}(U,\epsilon)=
2L- \langle N_{\rm tot} \rangle_{\rm gs}(U,-\epsilon-U).
\end{eqnarray}

In the strong coupling limit ,$J=\infty$,
the energies of $N_{\rm tot}=L-1,L,L+1$
are determined as $0,\epsilon,2\epsilon+U$.
In particular when $\epsilon =U=0$,
these three states are degenerated.
We define
such a point as the tricritical point $(U_c,\epsilon_c)$
({\it i.e.}, 
$(U_c,\epsilon_c)=(0,0)$ at $J=\infty$).
In general,
from the electron-hole symmetry
one can show that the tricritical point
$(U_c,\epsilon_c)$,
if it exists,
must be on the symmetric line $\epsilon_c=-U_c/2$.

\begin{figure}
\resizebox{9cm}{!}{\includegraphics{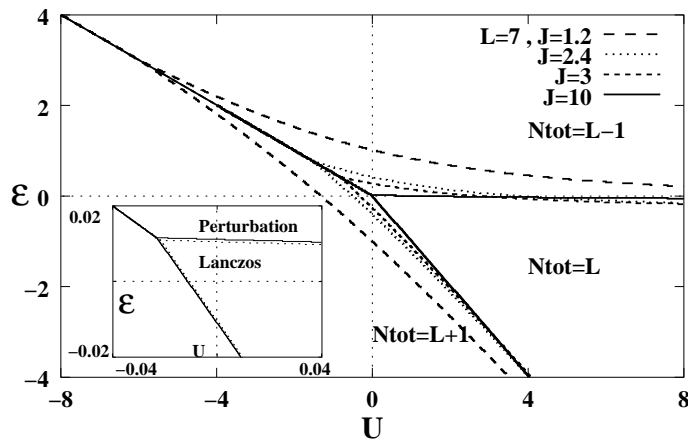}}
\caption{
Phase diagram of $\langle N_{\rm tot}\rangle _{\rm gs}(U,\epsilon)$
obtained from the Lanczos method using the same $L=7$ for $J=1.2, 2.4, 3, 10$.
Inset:comparison between the results by
the Lanczos method (dashed lines)
and 
the fourth order $t/J$ perturbation (solid lines)
for $J=10$.
}
\label{gsall}
\end{figure}

Figure~\ref{gsall} shows that
as decreasing $J$,
the domain of the doublet
$\langle N_{\rm tot}\rangle _{\rm gs}(U,\epsilon)=L$
({\it i.e.}, the magnetic state)
is extended. 
The transition lines of $J=10$ (the solid lines in Fig.~\ref{gsall})
are very close to those of the strong coupling limit.
In the strong coupling case,
$J/t \gg 1$,
one can use the perturbation theory
to estimate the boundaries.
In the inset of Fig.~\ref{gsall},
comparison is made between the fourth order perturbation results
and the numerical ones for $J=10$.
The agreement is satisfactory.

\begin{figure}
\resizebox{9cm}{!}{\includegraphics{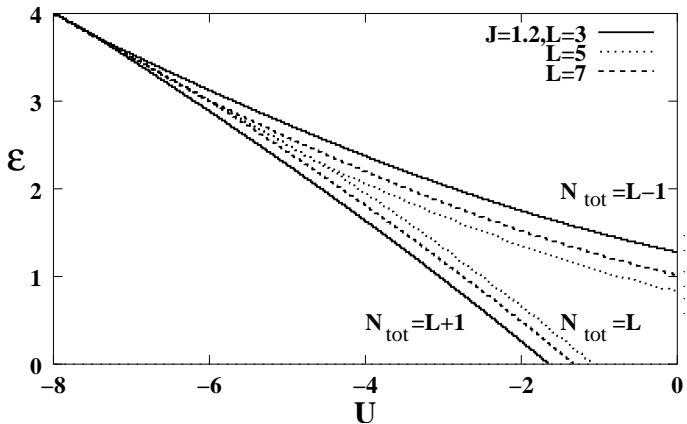}}
\caption{The finite size effect on the phase diagram obtained from
the Lanczos method for $J=1.2 , t=1$
with varying $L=3 , 5, 7$.
}
\label{gsJ12Lx}
\end{figure}

We comment on the finite-size effect on the phase diagram.
The numerical results shown in Fig.~\ref{gsall} are for a fixed system size,
$L=7$.
When $J\geq 2.4$,
the system-size dependence
of $\langle N_{\rm tot}\rangle _{\rm gs}(U,\epsilon)$
is very small and has reached a convergence within $L<7$.
But,
when $J$ becomes smaller
the system-size dependence
becomes more significant.
Fortunately, however,
the finite-size effect shows an alternating behavior
as shown in Fig.~\ref{gsJ12Lx}.
From the system-size dependence,
we may conclude that $-5.96< U_c< -4.54$ for $J=1.2$.
Table~\ref{tableUc} summarizes the results of
the tricritical points for various $J$,
where the errors are estimated from the finite-size effect.

\begin{table}
\begin{ruledtabular}
\begin{tabular}{cccc} 
& $J$ & $U_c$  \\ \hline
& 1.2 & -5.96 $ < U_c <$ -4.54  \\
& 2.4 & -1.58 $\pm $0.02 \\
& 3 & -0.82 $\pm $0.02  \\
& $\infty$ & 0
\end{tabular}
\end{ruledtabular}
\caption{Tricritical point $U_c(J)$ estimated from the Lanczos method.
$\epsilon_c$ is given by $-U_c/2$.}
\label{tableUc}
\end{table}

\section{Temperature dependence of susceptibilities}
In order to study the process of formation of the magnetic moment
as a function of temperature,
we calculate susceptibilities.
The values of susceptibilities at zero temperature depend on
the nature of the ground state.
At low temperatures,
susceptibilities diverge like a Curie-law if the ground state is
magnetic,
while converge to zero if the ground state is non-magnetic.

\subsubsection{definition of susceptibilities}
With a global magnetic field $h$,
we define total and impurity susceptibilities in the following way,
\begin{eqnarray}
\chi_{\rm tot}
&:=&{\partial \langle S^z_{\rm tot}\rangle \over \partial h},
\label{eqchitot}
\\
\chi_{\rm imp}
&:=&{\partial \langle s^z_{n_d}\rangle \over \partial h}
\label{eqchiimp}
,
\end{eqnarray}
where the z-component of the total spin is given by
$S^{z}_{\rm tot}:=\left(\sum_{n\neq n_d} S^{z}_n+\sum_{n} s^{z}_n\right)$.
The system consists of 
the $L-1$ sites of the Kondo insulator and one impurity
site.
Therefore,
in order to extract the impurity effect,
we define another impurity susceptibility:
\begin{eqnarray}
\chi_{d}
&:=&\chi_{\rm tot} - {L-1\over L} \chi_{\rm tot}^{\rm KL}
\label{eqchid}
,
\end{eqnarray}
where $\chi_{\rm tot}^{\rm KL}$ is defined by
$\chi_{\rm tot}$ of the 1D Kondo insulator
without any impurity (Eq.~\ref{eqHKL}).
Of course,
the contributions to this susceptibility from the Kondo insulator
parts are zero at zero temperature,
because both of them have the 
singlet ground states.

These susceptibilities are calculated by using
the finite-temperature
density-matrix renormalization group (finite-$T$ DMRG)~\cite{FTDMRG} method
as follows.
From
the maximum eigenvalue $\lambda_M$ of
the quantum transfer matrix ${\cal T}_M$,
the partition function $Z_{\rm KL}$ of the 1D Kondo insulator
without any impurity (Eq.~\ref{eqHKL})
is given by
\begin{eqnarray}
Z_{\rm KL}&=&\mbox{tr} {e^{-\beta H_{\rm KL}}}
=\lim_{M\rightarrow \infty} \mbox{tr} {{\cal T}_M}^{L/2}
\nonumber\\
&=& \lim_{M\rightarrow \infty} \lambda_M^{L/2}
, \mbox{ when } L\rightarrow \infty
,
\end{eqnarray}
where $M$ is the Trotter number.
The last equality is obtained in
the thermodynamic limit $L\rightarrow
\infty$.
Following S.Rommer~\cite{ROMMER},
we obtain the partition function $Z$ of the model Hamiltonian (Eq.\ref{eqHm})
by the quantum transfer matrix of the impurity site
${\cal T}_M^{\rm imp}$.
\begin{eqnarray}
Z
&=&\mbox{tr} {e^{-\beta H}}
=\lim_{M\rightarrow \infty} \mbox{tr} {{\cal T}_M}^{L/2-1}{{\cal T}^{\rm imp}_M}
\nonumber\\
&=&\lim_{M\rightarrow \infty} \lambda_M^{L/2-1}
\langle \Psi_M^{\rm L}| {{\cal T}^{\mbox{\rm imp}}_M}|\Psi_M^{\rm R}\rangle 
, \mbox{ when } L\rightarrow \infty
,
\end{eqnarray}
where $\lambda_M$ and
the normalized eigenvectors
$\langle \Psi_M^{\rm L}|,|\Psi_M^{\rm R}\rangle $ 
are obtained from the finite-$T$ DMRG for a given trotter number M.
The susceptibility is obtained from the free energy $F$ or the magnetization
$m$ numerically as $-\delta F / {\delta h^2\over 2}$ or $\delta m /\delta h$
under a small magnetic field $\delta h$.
Typical truncation errors in the infinite-system algorithm of
finite-$T$ DMRG method are $10^{-2}$ at the lowest
temperature $kT=0.12$ for the Trotter number $M=200$.

In addition,
non-local susceptibilities can be defined
by applying a small magnetic field
on a site,
for example at the impurity site,
and measure magnetization at different sites.
We will use such susceptibilities
in latter discussions
to investigate how effects of the impurity moment extend to
the bulk of the Kondo insulators.

\subsubsection{Result of susceptibilities}
In the strong coupling limit,
$J=\infty$,
it is simple to calculate the susceptibilities,
because the magnetic moment is completely localized at the impurity site:
$\chi^{(0)}_{\rm tot}=(L-1)\beta{1+2e^{-J\beta/3}\over e^{3J\beta/4}+4+e^{-J\beta/3}}
+\chi^{(0)}_{\rm imp}$
,
$\chi^{(0)}_{\rm imp}=\chi^{(0)}_{d}=\beta{1\over 2(2+e^{\beta \epsilon}+e^{-\beta(\epsilon+U)})}$
.
These susceptibilities at low temperatures
show
a sharp transition
at the boundaries around the magnetic solution.
In the magnetic region,
the ratio of 
the susceptibilities to the Curie law susceptibility
($\chi_{\rm C}=1/4 kT$)
converge to unity in the limit of $T\rightarrow0$:
$\chi^{(0)}_{\rm imp}/\chi_{\rm C}=\chi^{(0)}_{d}/\chi_{\rm C}=1$.

At high temperatures $T \gg J$ and $t$ ,
susceptibilities of any $J/t$
show good agreement with the susceptibilities of the strong coupling limit,
since
correlations between different sites are negligible.
In fact,
the high-temperature expansion of the susceptibility $\chi$,
which is obtained from the second order perturbation of $t/J$,
is given by
$\chi_{\rm imp}-\chi^{(0)}_{\rm imp}=- t^2 \beta^3 /4 $
.

\begin{figure}
\resizebox{9cm}{!}{\includegraphics{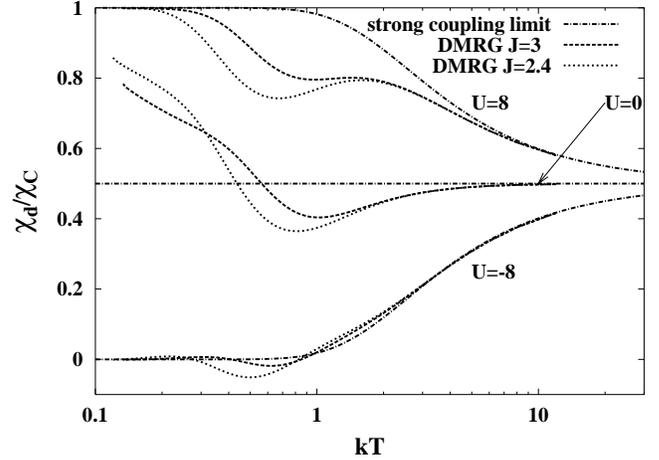}}
\caption{Ratio of the susceptibility $\chi_{d}$
to the
Curie susceptibility $1/4kT$ calculated
by the finite-$T$ DMRG method
for $J=3, 2.4$ and for $\epsilon=-U/2$;
$U=8, 0, -8$.
}
\label{dkaiFpaper}
\end{figure}

On the other hand,
at low temperatures
the values of susceptibilities 
depend on the ground state as we mentioned
at the beginning of this section.
Figure \ref{dkaiFpaper} shows $\chi_{d}/\chi_{\rm C}$ for various $J$
and $(U,\epsilon)$ on the symmetric line:
$U=8 , 0 , -8$ and $\epsilon=-U/2$.
First we note that
$\chi_d / \chi_{\rm C}$ for $U=0$ and $J=\infty$
is special and temperature independent.
This is due to the fact that
$U=0$ and $\epsilon=0$ is
the tricritical point
in the strong coupling limit.
Except for $U=0$ and $\epsilon=0$ case,
all lines are pointing towards either zero or unity
in the zero temperature limit.
These behaviors are consistent with the values
of the tricritical point in Table~\ref{tableUc}
estimated from the ground state.
However the temperature dependence of $\chi_d / \chi_{\rm C}$ is
quite different for $U=8$ and 0.

\begin{figure}
\resizebox{9cm}{!}{\includegraphics{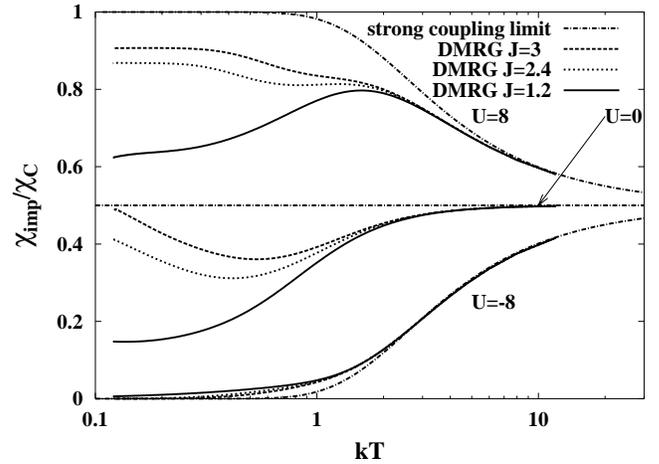}}
\caption{Ratio of the susceptibility $\chi_{\rm imp}$
to the
Curie susceptibility $1/4kT$ calculated
by
the finite-$T$ DMRG method
for $J=3, 2.4, 1.2$ and for $\epsilon=-U/2$;
$U=8, 0, -8$.
}
\label{kaiFpaper}
\end{figure}

To understand the difference,
we plot $\chi_{\rm imp} / \chi_{\rm C}$ in Fig.~\ref{kaiFpaper}.
One can see that $\chi_{\rm imp} / \chi_{\rm C}$ approaches to a finite value
for $U=8$ and 0 while it goes to zero for $U=-8$.
Furthermore the finite value seems to depend on $U$ and$J$.
As can be seen from the definition of $\chi_{\rm imp}$
(Eq.~\ref{eqchiimp}),
only the induced moment at the impurity site is measured
for this susceptibility,
while the entire moment induced in the system is measured for $\chi_d$
(Eq.~\ref{eqchid}).
Therefore the different limit $\chi_{\rm imp} / \chi_{\rm C}$ may be
understood as due to the difference in the spatial spread
of the induced magnetic moment.
Naturally,
we expect that the induced moment has a larger amplitude
at the impurity site for larger $U$ ($>U_c$) and a wider
spread for smaller $J$ since the correlation length
of the Kondo insulator gets longer as $J$ is decreased~\cite{RKKYint}.
We may conclude that
the limiting value of $\chi_{\rm imp} / \chi_{\rm C}$ is larger for larger $U$
($>U_c$) and larger $J$.
This is exactly what we observe in Fig.~\ref{kaiFpaper}


In order to investigate spatial spread of the magnetic moment,
we calculate the susceptibilities
\begin{eqnarray}
\chi_{s_n}
&:=&
{\delta \langle s^z_n \rangle  \over  \delta h_d}
,
\end{eqnarray}
where $h_d$ is the local magnetic field applied on the impurity site only.

When a magnetic field is applied on one site in the Kondo lattice model,
the induced magnetic moments show
the RKKY oscillatory behavior
\begin{eqnarray}
\langle s_n \rangle
&\propto&
(-1)^{(n-n_0)} \exp( -| n-n_0 | / \xi )
,
\end{eqnarray}
where $\xi$ is the correlation length.
Thus we expect that the
non-local susceptibilities $\chi_{s_n}$
also show a similar behavior.

\begin{figure}
\resizebox{9cm}{!}{\includegraphics{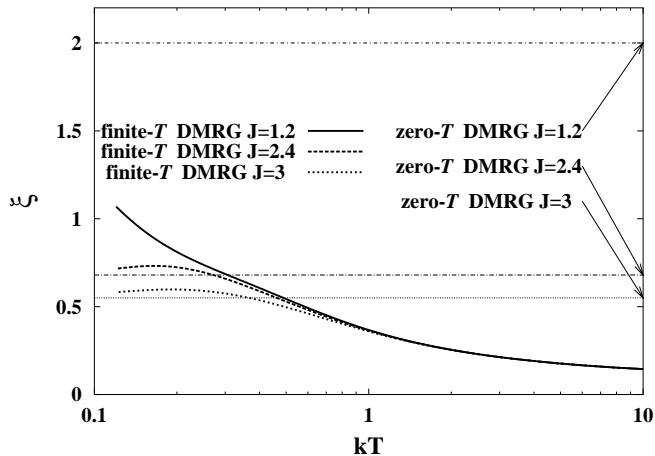}}
\caption{The correlation length $\xi$ estimated
from $\chi_{s_{n_d}}$ and $\chi_{s_{n_d+1}}$
for $J=1.2, 2.4, 3$.
The correlation lengths
of the bulk Kondo insulator obtained by the zero-temperature DMRG method
[13] 
are also shown.
}
\label{RKKYcor}
\end{figure}

It is interesting to estimate the correlation length from
the ratio between $\chi_{s_{n_d+1}}$ and $\chi_{s_{n_d}}$
through the relation
\begin{eqnarray}
\xi^{-1} &=&
- \log(-\chi_{s_{n_d+1}} /\chi_{s_{n_d}})
.
\end{eqnarray}
Figure \ref{RKKYcor} shows that 
values of $\xi$ at low temperatures
estimated from the above relation
are close to the correlation length of the 1D Kondo insulator
evaluated at $T=0$ by the zero-temperature DMRG~\cite{RKKYint}.
The value of $\xi$
increases
with decreasing $J$,
which is consistent with the
divergence of
the correlation length at $J=0$~\cite{RKKYint}.

Finally we would like to comment on the non-monotonic temperature
dependence observed in Fig. 4.  The decrease of the ratio of the 
susceptibility to the Curie susceptibility
around $T=2$ with lowering temperature is due to formation of
conduction band. This is a general trend which is shown
by the above second order perturbation theory with respect to $t/J$.
On the other hand, the upturn around $T=0.5 \sim 0.8$ may be traced
back to the effect of formation of the Kondo singlet states on both
sides of the impurity.
In fact, this idea is consistent with the 
fact that the minimum occurs around the spin gap temperature of the
Kondo insulators,
however,
further investigation is necessary to draw a
definitive conclusion.

\section{Conclusions}
In this paper we have studied the problem of formation of
magnetic moment when a Kondo insulator is doped with an 
impurity without the localized spin.
One dimensional Kondo 
lattice or Kondo chain is treated in this article.
The ground state
phase diagram is obtained by the Lanczos method.
Temperature
dependence of the impurity susceptibilities calculated by the
finite-$T$ DMRG are consistent with the phase diagram.

In contrast to the naive expectation,
the region of stable magnetic 
moment in the $(U, \epsilon)$ plane is wider for smaller $J$.
It is surprising that the region of stable magnetic moment extends
to the region of negative $U$ for a finite $J$.
The key to 
understand this result is the extension of the formed magnetic 
moment.
Longer correlation length of the Kondo insulator with 
smaller $J$ means larger spread of the magnetic moment.
It means that
the effect of Coulomb repulsion
(attraction for negative $U$)
is weakened for smaller $J$.
In the weak coupling case, 
the correlation length diverges exponentially as a function of $J$.
On the other hand,
the charge gap which is linear in $J$ \cite{SHIBATA}
acts in the Kondo lattice parts
as a positive effective potential.

It is interesting to compare the effects of depletion of localized 
spins between the two-leg spin ladder and the one dimensional
Kondo insulator.
The both system have spin gaps as a common feature.
For the spin ladder,
depletion of a spin always leads to formation of a
localized moment around the depleted site.
For the Kondo chain, 
formation of a magnetic moment depends on $(U, \epsilon)$.

A finite concentration of depletion may lead to quasi long
range order characterized by gap-less spin excitations in the spin
ladder.
On the other hand for the Kondo insulator,
one can speculate that there may be two cases.
In one case,
the ground state may be non-magnetic,
while in the other case there may be quasi long range
anti-ferromagnetic ordering.
This will be an interesting future problem.

Finally let us consider the effect of non-magnetic impurities in 
real three dimensional Kondo insulators.
We believe that the first conclusion
that the formation of magnetic moment is generally
easier in Kondo insulators than in ordinary metals is also valid in higher 
dimensions.
It is also probable to consider that there may be the two cases,
non-magnetic and magnetic ground states for finite concentrations
of non-magnetic impurities in the Kondo insulators
in realistic three dimension.
However,
estimation of the critical concentration for each case 
needs more elaborate investigation and clearly beyond the range 
of the present paper. 

\acknowledgments
This work is supported by Grant-in-Aid for Scientific Research Areas (B)
from the Ministry of Education,
Science,
Sports,
Culture and Technology and
by New Energy and Industrial Technology Development Organization (NEDO).

\end{document}